\documentclass[12pt,preprint]{emulateapj}
\usepackage{graphicx}
\usepackage{wasysym}

\shorttitle{A Stellar Companion to 30 Arietis B}
\shortauthors{Stephen R. Kane et al.}
\slugcomment{Submitted for publication in the Astrophysical Journal}

\begin{document}

\title{On the Stellar Companion to the Exoplanet Hosting Star 30 Arietis B}
\author{
  Stephen R. Kane\altaffilmark{1},
  Thomas Barclay\altaffilmark{2,3},
  Michael Hartmann\altaffilmark{4},
  Artie P. Hatzes\altaffilmark{4},
  Eric L.N. Jensen\altaffilmark{5},
  David R. Ciardi\altaffilmark{6},
  Daniel Huber\altaffilmark{7,8},
  Jason T. Wright\altaffilmark{9,10},
  Elisa V. Quintana\altaffilmark{2}
}
\email{skane@sfsu.edu}
\altaffiltext{1}{Department of Physics \& Astronomy, San Francisco
  State University, 1600 Holloway Avenue, San Francisco, CA 94132,
  USA}
\altaffiltext{2}{NASA Ames Research Center, M/S 244-30, Moffett Field,
  CA 94035, USA}
\altaffiltext{3}{Bay Area Environmental Research Institute, 596 1st
  Street West, Sonoma, CA 95476, USA}
\altaffiltext{4}{Th\"uringer Landessternwarte, D-07778 Tautenburg,
  Germany}
\altaffiltext{5}{Dept of Physics \& Astronomy, Swarthmore College,
  Swarthmore, PA 19081, USA}
\altaffiltext{6}{NASA Exoplanet Science Institute, Caltech, MS 100-22,
  770 South Wilson Avenue, Pasadena, CA 91125, USA}
\altaffiltext{7}{Sydney Institute for Astronomy (SIfA), School of
  Physics, University of Sydney, NSW 2006, Australia}
\altaffiltext{8}{SETI Institute, 189 Bernardo Avenue, Mountain View,
  CA 94043, USA}
\altaffiltext{9}{Department of Astronomy and Astrophysics,
  Pennsylvania State University, 525 Davey Laboratory, University
  Park, PA 16802, USA}
\altaffiltext{10}{Center for Exoplanets \& Habitable Worlds,
  Pennsylvania State University, 525 Davey Laboratory, University
  Park, PA 16802, USA}


\begin{abstract}

A crucial aspect of understanding planet formation is determining the
binarity of the host stars. Results from radial velocity surveys and
the follow-up of {\it Kepler} exoplanet candidates have demonstrated
that stellar binarity certainly does not exclude the presence of
planets in stable orbits and the configuration may in fact be
relatively common. Here we present new results for the 30 Arietis
system which confirms that the B component hosts both planetary and
stellar companions. Keck AO imaging provides direct detection of the
stellar companion and additional radial velocity data are consistent
with an orbiting star. We present a revised orbit of the known planet
along with photometry during predicted transit times. Finally, we
provide constraints on the properties of the stellar companion based
on orbital stability considerations.

\end{abstract}



\section{Introduction}
\label{intro}

The binarity of stars is a topic of ongoing research, particularly in
light of the plethora of exoplanets discovered over the past couple of
decades. Exoplanets orbiting stars with a binary companions pose
significant implications for formation theories, such as orbital
stability \citep{hol99}, and the period-mass \citep{zuc02} and
period-eccentricity \citep{egg04} distributions. The searches for
stellar companions to the host stars of {\it Kepler} exoplanet
candidates has become an important component of the candidate
validation process \citep{dre14,eve15,wan14}. Attempts to detect
binarity for the brightest exoplanet host stars are also underway
\citep{cre12,cre13}, and are often used to place constraints on
additional planetary companions \citep{kan14}.

When it comes to multiplicity, one of the more exotic exoplanetary
systems is that of 30 Arietus (hereafter 30 Ari). 30 Ari is a bound
visual binary whose main components are both main sequence F stars
(F5V and F6V) separated by 38.1\arcsec (1,500~AU). The A and B
components are both relatively bright ($V$ magnitudes of 6.48 and 7.09
respectively). 30~Ari~A is a spectroscopic binary \citep{ada19,mor74}
with an orbital period of 1.1 days. 30~Ari~B (HD 16232, HIP 12184, HR
764) was discovered by \citet{gue09} to have a $\sim$10~$M_J$
companion with an orbital period of 335 days. The discovery was made
using radial velocity (RV) observations which are not easy to
undertake for such an early-type star, despite its brightness. The
reason for this is that the spectra of early-type stars have a
relatively small number of absorption lines and also tend to have
rapid rotation rates, thus inhibiting precision RV measurements. The
brightness of 30~Ari~B in close proximity to the equally bright A
component also proves problematic for photometric observations and so
the system remained relatively unobserved for the years following the
exoplanet discovery. Recently 30~Ari~B was revisited using the
adaptive optics capabilities of the Robo-AO system \citep{bar14} with
target selection from the FG-67 database \citep{tok14}. The survey
detected a stellar companion to the star \citep{rid15} that was
further described by \citet{rob15}.

Here we present new observations of the 30~Ari~B system that
independently confirm the presence of a stellar companion in addition
to the known planet orbiting the host star. Section \ref{stellar}
outlines the properties of 30~Ari~B relevant to the subsequent
analysis. Section \ref{detection} describes the detection of the
stellar companion from Keck observations and the likelihood of the
stars being bound. New RV and photometric data are presented in
Section \ref{rvphot} which both are used to support the detection of
the stellar companion and refine the properties of the known
planet. Constraints on the physical and orbital properties of the
stellar companion from these observations and orbital stability
considerations are described in Section \ref{properties}. We provide
concluding remarks in Section \ref{con} including a discussion of
names for the system components.


\section{Host Star Properties}
\label{stellar}

\begin{figure*}
  \begin{center}
    \begin{tabular}{cc}
      \includegraphics[width=8.2cm]{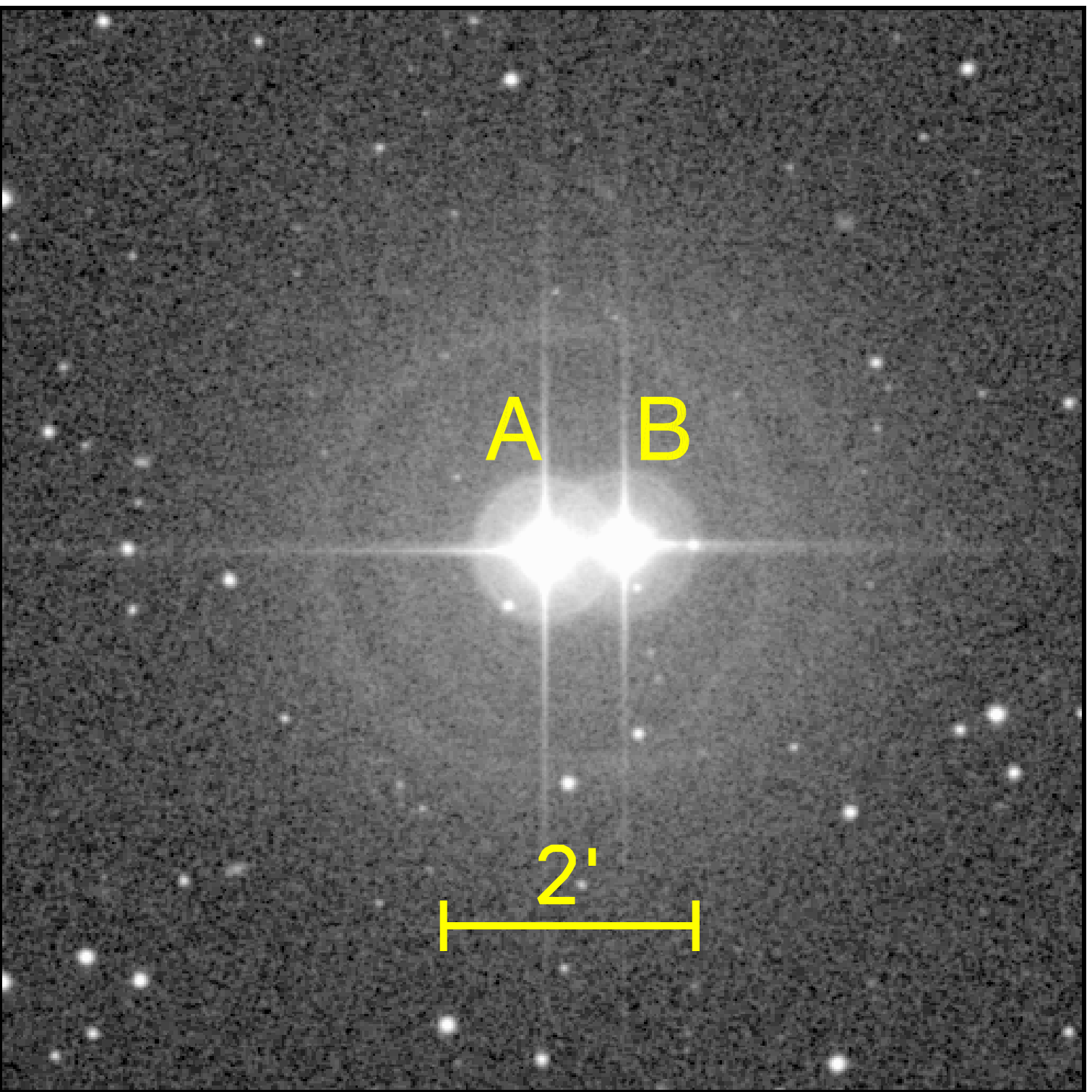} &
      \includegraphics[width=8.2cm]{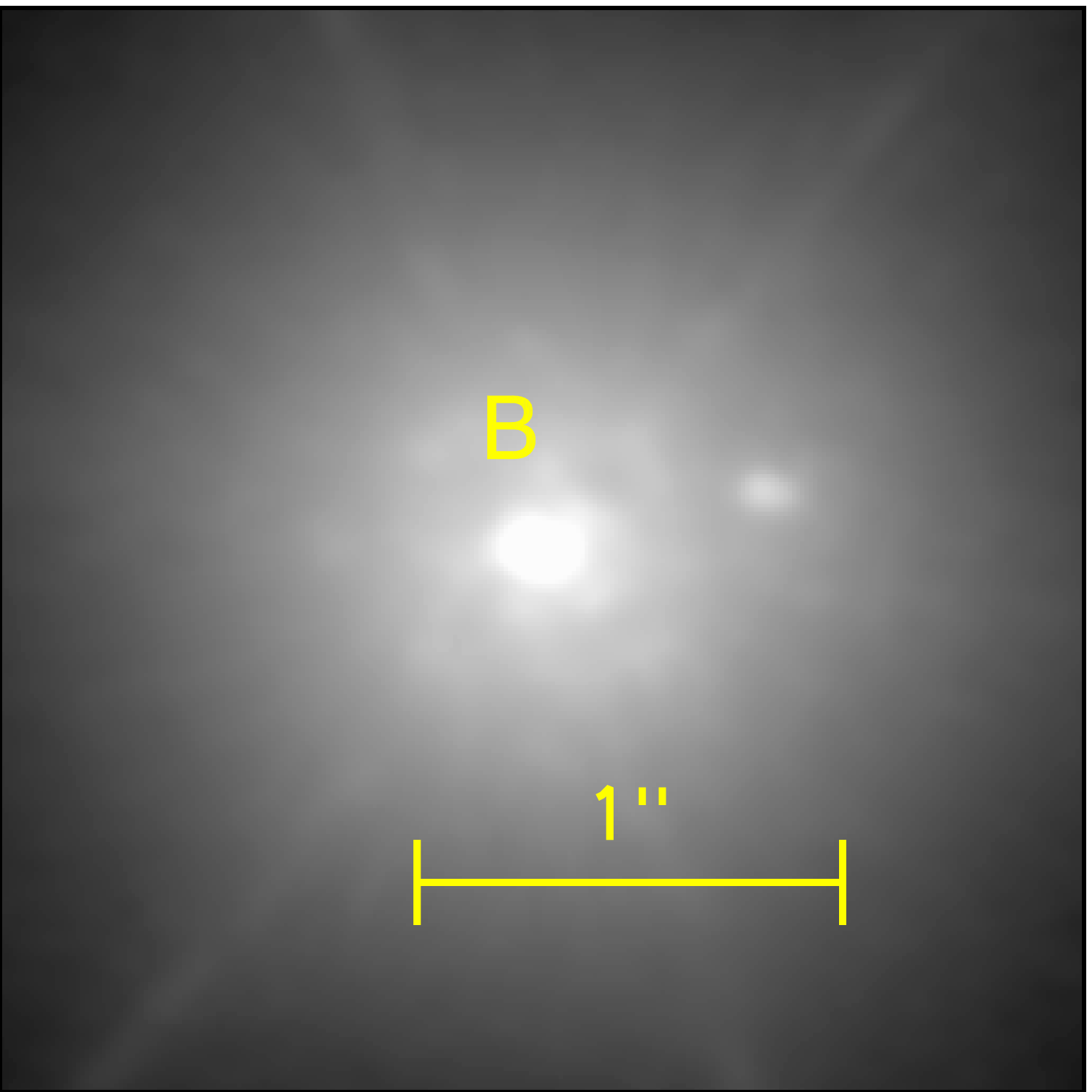}
    \end{tabular}
  \end{center}
  \caption{Images of the 30 Ari system from the Digital Sky Survey
    (Left) and Keck observations (right). In both cases the field
    orientation is up-North and left-East. Left: Image of the 30 Ari
    system centered on the A component. Right: Keck/NIRC2 combined
    image of 30 Ari B showing the presence of the stellar companion.}
  \label{imagefig}
\end{figure*}

This paper compiles imaging, RV, and photometric data of the 30~Ari
system. The advantage of combining these datasets is to maximize
constraints on both the kinematic and intrinsic luminosities of the
component members. Determining these properties of the individual
components depends heavily on the properties of 30~Ari~B. The
fundamental stellar properties of the star have been published
numerous times in the literature, most recently by \citet{tsa14}. In
order to compile a self-consistent set of stellar parameters relevant
to this work, and comparison with previous work on the planetary
companion (see Section \ref{rv}), we adopt those from \citet{van07}
and \citet{gue09}, shown in Table \ref{stellartab}. A particular
reason for selecting these stellar parameters is to be consistent with
the previous RV measurements of \citet{gue09} such that a direct
comparison of the Keplerian orbital solutions may be made (see
Section~\ref{rvphot}).

\begin{deluxetable}{lc}
  \tablecaption{\label{stellartab} 30~Ari~B Stellar Parameters$^{(1)}$}
  \tablehead{
    \colhead{Parameter} &
    \colhead{Value}
  }
  \startdata
  $J$                            & 6.080 \\
  $V$                            & 7.091 \\
  $B-V$                          & 0.510 \\
  Proper motion ($\alpha, \delta$) (mas)$^{(2)}$ & 150.75, -12.79 \\
  Parallax (mas)$^{(2)}$                 & $24.52 \pm 0.68$ \\
  Distance (pc)$^{(2)}$          & $40.8 \pm 1.1$ \\
  $M_\star$ ($M_\odot$)          & $1.16 \pm 0.04$ \\
  $R_\star$ ($R_\odot$)          & $1.13 \pm 0.03$
  \enddata
  \tablenotetext{(1)}{\citet{gue09} and references therein.}
  \tablenotetext{(2)}{\citet{van07}}
\end{deluxetable}


\section{Detection of a Stellar Companion}
\label{detection}

\begin{figure}
  \includegraphics[width=8.2cm]{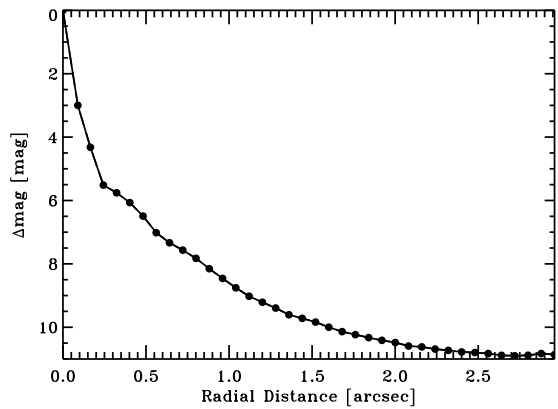}
  \caption{The 5$\sigma$ Keck image sensitivity (in units of $\Delta$
    magnitude) as a function of separation from the host star.}
  \label{deltamagfig}
\end{figure}

Shown in Figure \ref{imagefig} (left) is an $\sim$9\arcmin \ FOV image
of the 30 Ari visual binary extracted from the Digital Sky
Survey\footnote{\tt https://archive.stsci.edu/cgi-bin/dss\_form},
centered on the A component. Our observations of 30~Ari~B were
acquired using NIRC2 with the AO system at Keck during the night of
August 9th, 2014. We used the standard AO configuration for NIRC2
imaging observations, the details of which may be found in the NIRC2
Observer's Manual\footnote{\tt
  http://www2.keck.hawaii.edu/inst/nirc2/Manual/ObserversManual.html}. Sky
conditions were poor (thin cirrus clouds) but sufficient to complete
the observations given the brightness of the target. The camera was
used in the narrow camera mode with a $J$-band filter. A total of nine
0.2 second exposures were acquired and co-added to produce a combined
smoothed frame from which to conduct the analysis. The sensitivity of
the observations to fainter stellar companions is demonstrated in
Figure \ref{deltamagfig} which shows the 5$\sigma$ detection limit as
a function of radial separation from the host star.

\begin{figure}
  \includegraphics[width=8.2cm]{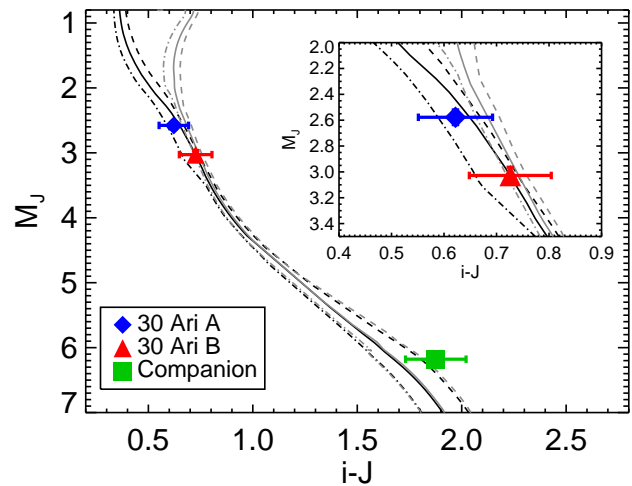}
  \caption{0.5 Gyr (black) and 1.5 Gyr (gray) Dartmouth stellar
    isochrones with $\mathrm{[Fe/H]} = -0.1$ (dash-dotted lines), 0.1
    (solid lines) and 0.3 (dashed lines). Note that the 0.5 Gyr
    isochrone has been obtained by linearly interpolating the original
    $<1$\,Gyr isochrone grid available in the Dartmouth database. The
    inset shows a zoom on the position of 30 Ari A and B.}
  \label{cmd}
\end{figure}

The combined Keck image is shown in Figure \ref{imagefig}
(right). 30~Ari~B is at the center of the frame and the stellar
companion is plainly visible to the right of the host
star. Measurements of the stellar profile centroids and the NIRC2
pixel scale (0.009942\arcsec/pixel) show that the stars are separated
by 0.536\arcsec. The uncertainty in the $X$-direction is 0.646 pixels,
equivalent to 0.00642\arcsec or 6.42~mas in RA. Similarly the
uncertainty in the $Y$-direction is 0.245 pixels, equivalent to
0.00244\arcsec or 2.44~mas in Dec. Thus the separation of the stars is
$0.536\arcsec \pm 0.007\arcsec$ with a position angle of $-73.6\degr
\pm 0.1\degr$ (east of north).

The relative photometry between the two stars was estimated into two
ways and the results compared. Because of the poor seeing on the night
of the observations, there was not a clean centralized peak of the
primary. To aid in the photometry, the final image was convolved with
a 2-D circularly symmetric gaussian with the full-width set to 4
pixels: approximately half the full-width of the measured PSF. The
first estimate utilized aperture photometry on each star where the
aperture radius was set to the half-width of the convolved PSF (5
pixels). The second estimate was performed by fitting a 2-D gaussian
to each of the stellar PSFs and subtracting the gaussians from the
image until the residuals were minimized. The total flux in the
gaussian PSFs were used to estimate the relative magnitudes of
stars. The final relative photometry was determined from an average of
the two methods, and the difference in the two methods was added in
quadrature to the formal statistical uncertainties in the aperture and
psf photometry. We find that the magnitude difference between the two
stars $\Delta_J = 3.15 \pm 0.07$. Using the distance estimate of Table
\ref{stellartab} leads to a projected separation of $21.9 \pm 0.7$~AU
and a companion absolute $J$ magnitude of $6.18 \pm 0.09$. This is
consistent with the companion being a late-type dwarf with an
approximate spectral type of M1-3 \citep{boy12}.

The issue remains as to whether the detected companion is indeed
gravitationally bound to the host star. No astrometric motion was
detected through the analysis of {\it Hipparcos} data by
\citet{ref11}. This null-detection is not surprising however
considering that the orbital period of the stellar companion is much
longer than the time baseline of the {\it Hipparcos} observations. The
proper motion of 30~Ari~B according to \citet{van07} is $0.151\arcsec
\pm 0.00075\arcsec$. The astrometric results of \citet{rob15} confirm
that the newly detected companion to 30~Ari~B has a common proper
motion, increasing the likelihood that they are bound. To investigate
this further, we adopt the statistical validation techniques described
\citet{hor14}. The 5$\sigma$ detection limit shown in Figure
\ref{deltamagfig} is similar to the detection limit achieved with the
Differential Speckle Survey Instrument (DSSI) on Gemini-North, shown
in Figure 9 of \citet{hor14}. Linear interpolation of the figure bins
indicates that the likelihood of our detected companion being bound to
30~Ari~B is $>82$\%. However, the observations of \citet{hor14} were
of the {\it Kepler} field which has a higher density of stars. To
account for that, we used the TRILEGAL code\footnote{\tt
  http://stev.oapd.inaf.it/cgi-bin/trilegal} \citep{gir05} to
determine the relative number of stars along the respective
lines-of-sight for 30~Ari~B and the {\it Kepler} field. A 1 square
degree search with the TRILEGAL model yields 16,210 line-of-sight
companions for 30~Ari~B and 167,936 line-of-sight companions for the
{\it Kepler} field ($l=76.53$, $b=13.29$). Assuming the binarity rate
does not change, the probability that the companion detected near
30~Ari~B is gravitationally bound is increased by the ratio of the
number of companions predicted, which is a factor of $\sim$10. Thus,
the probability that the detected companion is bound to 30~Ari~B is
$\sim$100\%.

At the point of submitting this work, we learned that the stellar
companion to 30~Ari~B has also been detected by the Robo-AO team
\citep{rid15}. We present the Keck AO component of these results as an
independent detection of this companion. Their observations were
conducted using an $i$ filter and reveal a similar angular separation
of 0.536\arcsec. If the detected companion is gravitationally bound,
the 30~Ari~B components should have colors and absolute magnitudes
that are compatible with stellar isochrones. To test this, we combine
the $J$-band detection from our Keck observations with the $i$-band
detection from Robo-AO to place the components on a color-magnitude
diagram.

The available $i$-band photometry of 30~Ari~B from the Sloan
\citep{ahn12} and APASS\footnote{http://www.aavso.org/apass} surveys
is heavily saturated, and therefore not reliable for this system. To
derive an approximate $i$ magnitude, we fit $J-K$ and $M_J$ to a grid
of Dartmouth isochrones \citep{dot08} assuming zero reddening, which
is justified by the relatively small distance ($\sim$41~pc) to the
system. The metallicity of the 30~Ari system is poorly constrained,
with values ranging from near solar metallicity from Stromgren
photometry \citep{cas11} to $\mathrm{[Fe/H]} \sim +0.27$ from various
spectroscopic studies \citep[see][and references
  therein]{gue09}. Adopting a metallicity prior of $\mathrm{[Fe/H]} =
0.1\pm0.2$ for the isochrone fit, which approximately corresponds to
the central value and spread of the quoted literature values, we
derive a synthetic absolute magnitude of $M_i=3.75 \pm 0.08$~mag for
30~Ari~B. Combining this with $\Delta J=3.15\pm0.07$~mag and $\Delta
i=4.2\pm0.1$~mag, the corresponding colors are $i-J=0.73\pm0.08$~mag
for 30~Ari~B and $i-J=1.88\pm0.13$~mag for the detected stellar
companion, respectively.

Figure \ref{cmd} compares the positions of 30~Ari~B and the detected
companion in a $M_{J}$, $i-J$ color-magnitude diagram to 0.5--1.5~Gyr
isochrones for a range of metallicities. 30~Ari~A is also shown, with
an $i-J$ color derived using the same procedure as described
above. The comparison shows that all three components have colors that
are consistent with a given distance modulus, and hence are compatible
with being in a gravitationally bound system.


\section{Radial Velocities and Photometry}
\label{rvphot}

Here we present new RV and photometric data of 30~Ari~B in support of
our observations of the stellar companion.


\subsection{Revised Planetary Parameters and Linear Trend}
\label{rv}

\begin{figure*}
  \begin{center}
    \includegraphics[width=15.5cm]{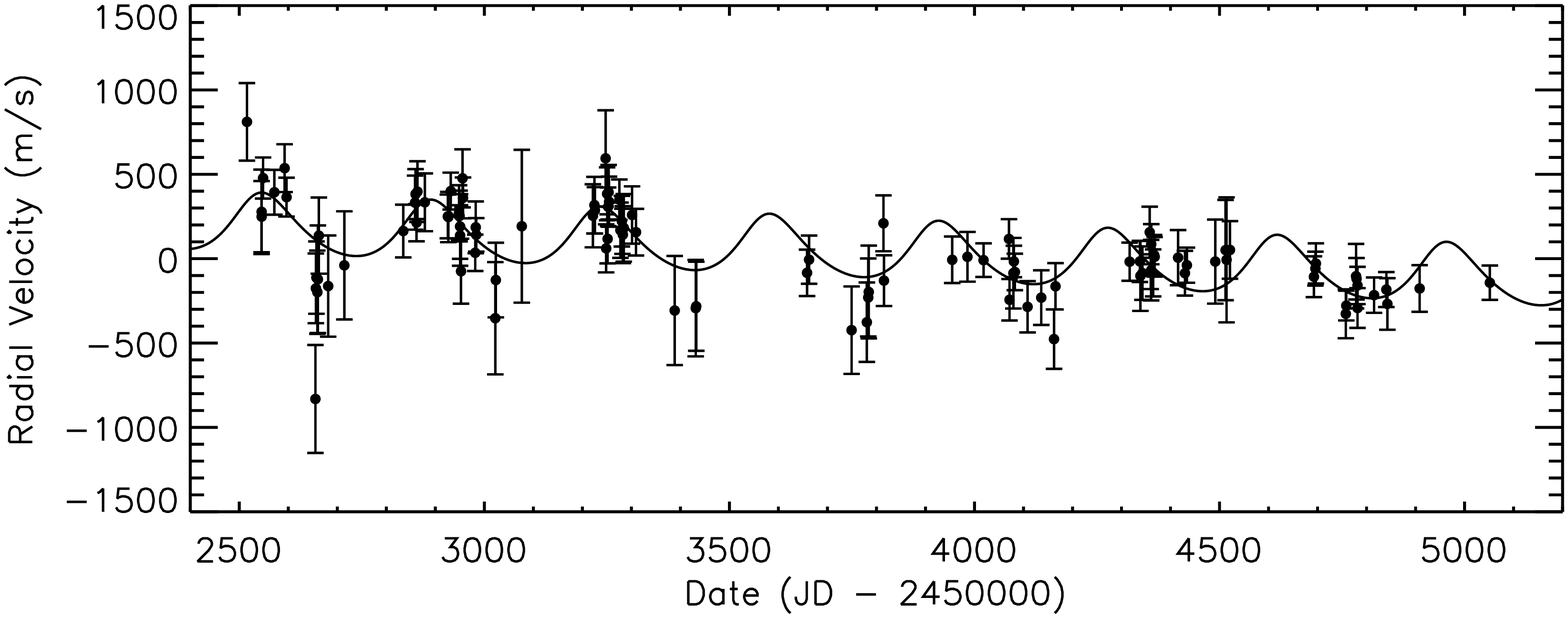} \\
    \includegraphics[width=15.5cm]{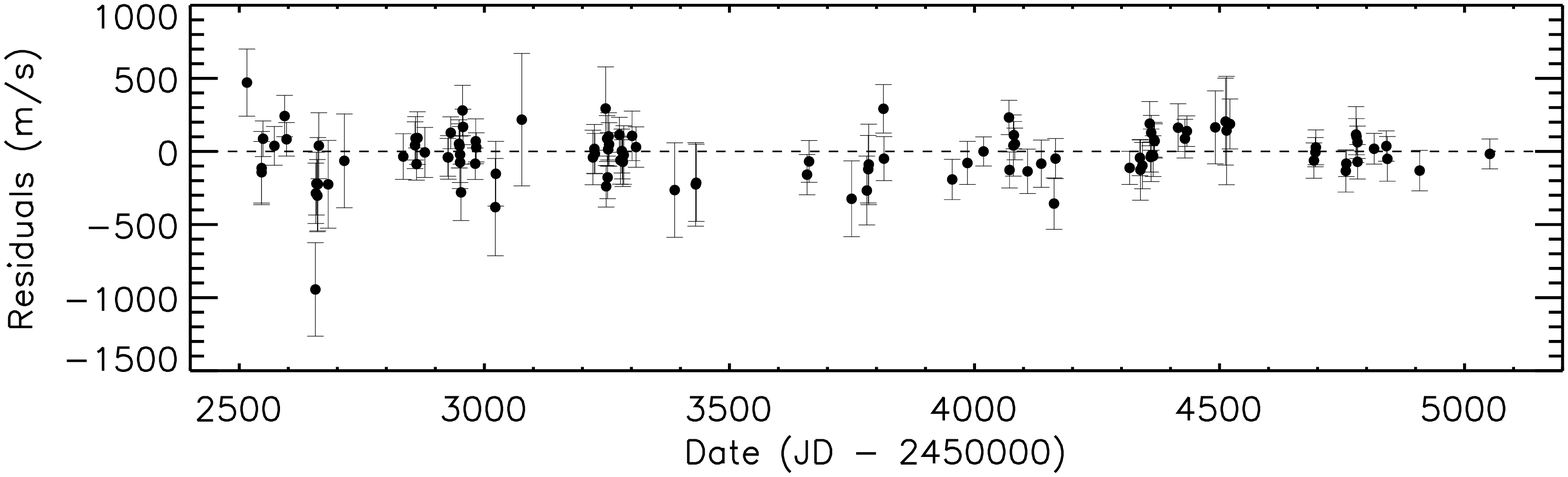}
  \end{center}
  \caption{All 110 RV measurements of 30~Ari~B acquired using the 2m
    Alfred Jensch telescope (see Table \ref{rvs}). Top: The new
    orbital solution including a linear trend due to the presence of
    the stellar companion, provided in Table \ref{planet}. Bottom: The
    RV residuals (observed minus computed) from the best-fit model.}
  \label{rvfig}
\end{figure*}

The RV dataset for 30~Ari~B published by \citet{gue09} consisted of 98
measurements and revealed the presence of a sub-stellar companion in a
335~day orbit around the host star. \citet{gue09} did not provide a
fit that included a linear trend free parameter since the presence of
such a trend was negligible in those data. Here we provide 12
additional measurements which extend the time baseline by $\sim$300
days and thus greater sensitivity to the possible influence of a
stellar companion. The data were acquired from continued observations
2m Alfred Jensch telescope of the Th\"uringer Landessternwarte
Tautenburg, described in detail by \citet{hat05}, and were reduced
with the same data pipeline as for those for \citet{gue09}. The data
were modeled using a partially linearized, least-squares fitting
procedure \citep{wri09}. Parameter uncertainties were estimated using
the BOOTTRAN bootstrapping routines developed by \citet{wan12}. The
new best-fit Keplerian orbital solution is shown in Table \ref{planet}
and Figure \ref{rvfig} along with the fit residuals. The parameters
include orbital period ($P$), time of periastron passage ($T_p$),
eccentricity ($e$), periastron argument ($\omega$), RV semi-amplitude
($K$), minimum planet mass ($M_p$), semi-major axis ($a$), and the RV
linear trend ($dv/dt$). The complete set of new and revised 110 RV
measurements are provided in Table \ref{rvs}.

\begin{deluxetable*}{lcc}
  \tablecaption{\label{planet} Keplerian Orbital Model}
  \tablewidth{0pt}
  \tablehead{
    \colhead{Parameter} &
    \colhead{Value \citep{gue09}} &
    \colhead{Value (This work)}
  }
  \startdata
\noalign{\vskip -3pt}
\sidehead{30 Ari B b}
~~~~$P$ (days)                    & $335.1 \pm 2.5$   & $345.4 \pm 3.8$  \\
~~~~$T_p$ (JD -- 2,440,000)       & $14538 \pm 20$    & $13222.1 \pm 42.4$ \\
~~~~$e$                           & $0.289 \pm 0.092$ & $0.18 \pm 0.11$ \\
~~~~$\omega$ (deg)                & $307 \pm 18$      & $337 \pm 57$ \\
~~~~$K$ (m\,s$^{-1}$)             & $272 \pm 24$      & $177 \pm 26$ \\
~~~~$M_p$\,sin\,$i$ ($M_J$)       & $9.88 \pm 0.94$   & $6.6 \pm 0.9$ \\
~~~~$a$ (AU)                      & $0.995 \pm 0.012$ & $1.01 \pm 0.01$ \\
~~~~$dv/dt$ (m/s/day)             & 0.0               & $-0.12 \pm 0.03$\\
\sidehead{System Properties}
~~~~$\gamma$ (m\,s$^{-1}$)        & - & $9.8 \pm 17.7$ \\
\sidehead{Measurements and Model}
~~~~$N_{\mathrm{obs}}$            & 98 & 110 \\
~~~~rms (m\,s$^{-1}$)             & 135 & 181.5 \\
~~~~$\chi^2_{\mathrm{red}}$       & -   & 0.82
  \enddata
\end{deluxetable*}

There are several notable changes over the orbital solution of
\citet{gue09}, shown in Table~\ref{planet}. The inclusion of a linear
trend is warranted by the extended baseline and the solution shows
that the trend is significant at the 4$\sigma$ level. The linear trend
has consequences for the Keplerian solution in that the orbital period
is slightly increased and the ``shape'' of the orbit (eccentricity and
periastron argument) is less well constrained since it is closer to
being circular. Another change of note is that the linear trend
partially compensates for the semi-amplitude of the RV variations
resulting in a smaller minimum mass for the sub-stellar companion of
6.6~$M_J$. The companion in question is thus more likely to be
planetary in nature than an object in the brown dwarf regime. Finally,
it should be noted that we have not excluded any of the significant RV
outliers (e.g., the measurement acquired at epoch 2,452,655.25, see
Table \ref{rvs}). Testing such exclusions did not significantly impact
the Keplerian orbital solution. The implications of the linear trend
for the detected stellar companion to 30~Ari~B are discussed in more
detail in Section \ref{massorbit}.

\LongTables
\begin{deluxetable}{cccc}
  \tablewidth{0pc}
  \tablecaption{\label{rvs} 30 Ari B Radial Velocities}
  \tablehead{
    \colhead{Date} &
    \colhead{RV} &
    \colhead{$\sigma$} \\
    \colhead{(JD -- 2,450,000)} &
    \colhead{(m\,s$^{-1}$)} &
    \colhead{(m\,s$^{-1}$)}
  }
  \startdata
 2515.600993 &  754.1050 &  229.87 \\
 2545.536478 &  220.9028 &  250.29 \\
 2545.544754 &  192.3126 &  210.82 \\
 2548.472892 &  421.2561 &  121.74 \\
 2571.576275 &  336.9965 &  132.71 \\
 2592.467015 &  479.7458 &  141.60 \\
 2596.415295 &  307.9014 &  114.94 \\
 2655.252792 & -888.2267 &  319.81 \\
 2656.242768 & -232.6306 &  206.70 \\
 2657.267185 & -168.8275 &  214.44 \\
 2659.260606 & -255.4376 &  246.96 \\
 2660.294048 & -178.7523 &  318.83 \\
 2662.268544 &   79.8795 &  225.79 \\
 2681.341841 & -219.0311 &  300.12 \\
 2714.276746 &  -96.3552 &  320.68 \\
 2834.523535 &  107.0504 &  156.15 \\
 2858.575149 &  275.5233 &  160.05 \\
 2859.600468 &  326.0843 &  148.02 \\
 2861.586390 &  156.6134 &  110.20 \\
 2863.588342 &  340.7115 &  180.12 \\
 2878.477610 &  277.3908 &  171.12 \\
 2925.422421 &  193.6456 &  129.42 \\
 2926.480403 &  190.0219 &  148.87 \\
 2931.400890 &  343.9288 &  109.36 \\
 2948.365052 &  213.9280 &  164.54 \\
 2949.384915 &  196.2251 &  150.77 \\
 2950.438364 &  134.0650 &  191.77 \\
 2950.444684 &   80.5700 &  179.43 \\
 2952.400062 & -131.2379 &  192.41 \\
 2955.422149 &  419.4916 &  172.12 \\
 2956.465067 &  303.5786 &  121.07 \\
 2981.367219 &  -21.7257 &  107.85 \\
 2982.388335 &  129.1056 &  153.39 \\
 2983.413199 &   84.5132 &   99.35 \\
 3022.290291 & -409.6576 &  332.74 \\
 3023.367630 & -183.1904 &  221.21 \\
 3076.292073 &  135.4815 &  453.15 \\
 3221.496697 &  197.4571 &  186.82 \\
 3224.547386 &  260.7274 &  167.63 \\
 3225.512468 &  229.8522 &  137.80 \\
 3247.392319 &  537.8008 &  284.92 \\
 3248.490970 &    4.2384 &  142.16 \\
 3250.480289 &  327.6601 &  156.85 \\
 3251.466370 &   60.7189 &  145.38 \\
 3252.492894 &  250.7651 &  114.60 \\
 3253.432338 &  337.5163 &  161.58 \\
 3254.450110 &  280.9421 &  148.70 \\
 3275.513744 &  292.7484 &  119.60 \\
 3277.546322 &  112.2383 &  164.09 \\
 3280.508460 &  165.4369 &  150.31 \\
 3281.621507 &  105.3440 &  133.67 \\
 3282.493267 &   85.4409 &  168.59 \\
 3284.350643 &  126.1022 &  200.37 \\
 3301.371752 &  202.2588 &  169.87 \\
 3309.444008 &  100.2754 &  138.43 \\
 3388.344614 & -364.0933 &  323.23 \\
 3431.291948 & -349.9726 &  285.70 \\
 3432.265935 & -338.8958 &  263.92 \\
 3658.364855 & -140.6013 &  137.69 \\
 3662.535227 &  -62.5995 &  142.97 \\
 3749.258017 & -480.4116 &  259.14 \\
 3780.366708 & -433.3393 &  235.41 \\
 3783.265338 & -286.6540 &  231.20 \\
 3784.288850 & -254.0487 &  275.03 \\
 3814.276631 &  152.9526 &  165.55 \\
 3815.290494 & -186.8709 &  150.55 \\
 3954.589497 &  -63.4947 &  137.82 \\
 3985.625596 &  -45.7626 &  147.38 \\
 4018.454970 &  -65.4861 &   99.70 \\
 4070.394848 &   60.2101 &  117.30 \\
 4071.515854 & -300.3802 &  122.88 \\
 4079.374428 & -142.8664 &  209.14 \\
 4080.365276 &  -73.1703 &   93.38 \\
 4082.438154 & -135.5155 &  108.22 \\
 4108.340755 & -342.0903 &  151.94 \\
 4136.254291 & -287.5546 &  161.89 \\
 4162.258663 & -533.4807 &  176.15 \\
 4165.348658 & -220.5430 &  137.41 \\
 4316.580289 &  -74.6582 &  113.25 \\
 4337.564193 &  -73.0027 &  121.86 \\
 4338.559366 & -158.5779 &  207.82 \\
 4342.615233 & -142.4315 &  158.33 \\
 4357.560127 &  100.1291 &  151.16 \\
 4359.572735 & -133.7737 &  103.00 \\
 4360.541338 & -124.8448 &  179.69 \\
 4360.576121 &   29.2391 &  118.99 \\
 4364.560715 & -141.0313 &  139.21 \\
 4366.536859 &  -34.9721 &  118.68 \\
 4367.554451 &  -45.7453 &  117.28 \\
 4415.427584 &  -50.7634 &  163.37 \\
 4429.384583 & -142.7605 &  132.56 \\
 4433.367364 &  -95.2091 &  104.31 \\
 4491.292838 &  -73.9298 &  249.26 \\
 4512.358374 &   -5.8152 &  297.29 \\
 4514.353088 &  -63.6416 &  370.58 \\
 4521.285415 &   -4.7234 &  170.85 \\
 4692.598912 & -163.7491 &  120.98 \\
 4695.601537 & -115.6291 &  100.32 \\
 4696.539080 &  -85.1723 &  119.14 \\
 4757.636700 & -383.5377 &  144.69 \\
 4758.610225 & -334.8741 &   87.91 \\
 4778.588013 & -160.5203 &  191.33 \\
 4779.531238 & -175.7052 &  123.25 \\
 4781.531450 & -215.3962 &  110.56 \\
 4781.573376 & -349.1836 &  118.00 \\
 4815.330030 & -272.7974 &  104.98 \\
 4840.347284 & -239.8999 &  103.61 \\
 4842.460148 & -325.0461 &  152.85 \\
 4908.311715 & -233.2704 &  138.19 \\
 5051.539182 & -199.2144 &  101.76
  \enddata
\end{deluxetable}


\begin{figure*}
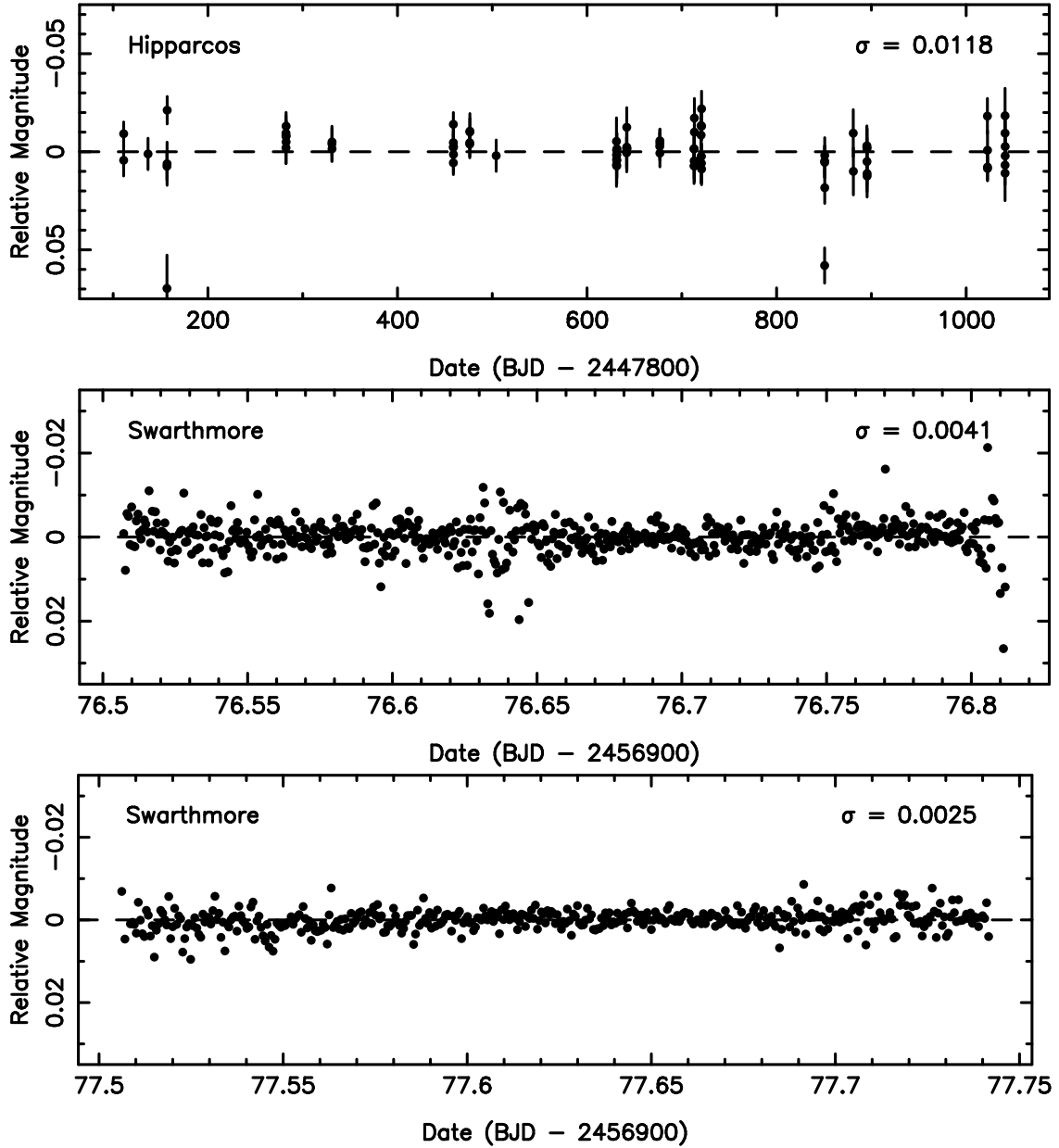

  \begin{center}
    \includegraphics[angle=270,width=15.0cm]{f05a.ps} \\
    \includegraphics[angle=270,width=15.0cm]{f05b.ps} \\
    \includegraphics[angle=270,width=15.0cm]{f05c.ps}
  \end{center}
  \caption{Photometry of 30~Ari~B from the {\it Hipparcos} mission
    (top panel) and the Swarthmore 0.6m telescope (middle and bottom
    panels). The number in the top-right of each panel is the
    1$\sigma$ rms scatter of the data points. Though no transit events
    were observed in the Swarthmore data, the star was found to be
    photometrically stable within a couple of millimags. Note the
    different vertical axis scales between the {\it Hipparcos} and
    Swarthmore plots.}
  \label{photfig}
\end{figure*}

\subsection{Potential Planetary Transit}
\label{phot}

With the detection of the low-mass stellar companion to 30~Ari~B, we
undertook the task of acquiring photometry that may have indications
of stellar variability. Variability studies of {\it Kepler} stars have
shown that F-type stars tend to have much shorter variation periods,
likely due to pulsations rather than the activity typical of low-mass
stars \citep{cia11,mcq12}.

The first photometric data source we examined was the photometry from
the {\it Hipparcos} mission, shown in the top panel of Figure
\ref{photfig}. These data demonstrate photometric stability at the
$\sim$1\% level. However, there are two significant outliers in the
photometry indicating an $\sim$5\% reduction in brightness of the host
star. The most intriguing aspect of these two outliers is that they
are separated by $\sim$693.8 days - approximately twice the revised
orbital period of the planet (see Section \ref{rv}). If such variation
were indeed due to the passage of the planet across the stellar disk,
the depth appears to be too large. Additionally, the probability of
the transit being detected in the sparsely sampled {\it Hipparcos}
data is extremely low. Nevertheless, to investigate this further, we
constructed a transit ephemeris based upon the {\it Hipparcos}
photometry since these data yield greater timing precision than
predictions based upon the RV data described in Section \ref{rv}.

\begin{figure*}
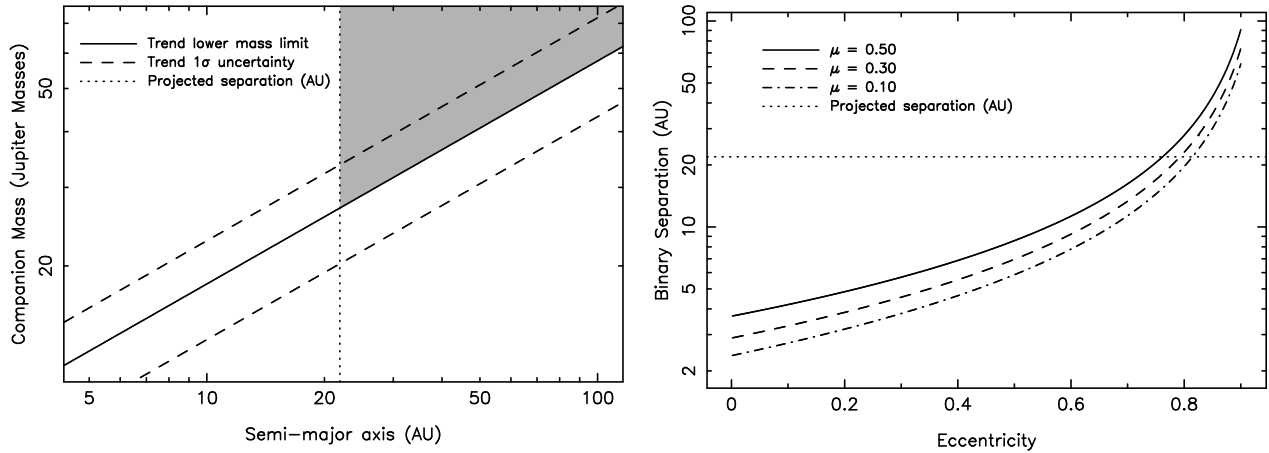

  \begin{center}
    \begin{tabular}{cc}
      \includegraphics[angle=270,width=8.2cm]{f06a.ps} &
      \includegraphics[angle=270,width=8.2cm]{f06b.ps}
    \end{tabular}
  \end{center}
  \caption{Limits on the properties of the companion star based on RV
    variations and stability constraints. Left: The lower mass limit
    for the stellar companion (solid line) imposed by the magnitude of
    the linear trend (assuming a circular orbit). The dashed lines are
    the 1$\sigma$ uncertainties and the vertical dotted line is the
    minimum projected separation (see Section \ref{detection}). The
    shaded area is the valid region based on these constraints. Right:
    The minimum separation of the 30~Ari~B binary components as a
    function of their orbital eccentricity that will allow the known
    planet to remain in a stable orbit. The horizontal dotted line
    represents the minimum projected separation of the stellar
    companion and the mass ratio is $\mu \sim 0.3$. Thus, the valid
    region of the plot is above both the dotted and dashed lines.}
  \label{propfig}
\end{figure*}

30~Ari~B was observed using the 0.6m telescope at the Peter van de
Kamp Observatory, Swarthmore College on the nights of November 15th
and 16th, 2014. Observations were conducted in good weather conditions
using an r' filter and 10 second exposures. 30~Ari~A provided a
natural comparison star from which to perform relative photometry
since it is similar in both brightness and color. Based upon the {\it
  Hipparcos} dips, a possible event was predicted for a JD of around
2,456,976. These data are shown in the middle and bottom panels of
Figure \ref{photfig}. Though no event similar to that seen in the {\it
  Hipparcos} data was detected, the star was observed to be
consistently stable at the level of a couple of millimags. These data
also rule out significant stellar pulsations of periods less than
$\sim$6~hours. It is certainly possible that the outlier measurements
in the {\it Hipparcos} dataset are simply spurious, but the curious
coincidence with the orbital period of the planet leads us to
encourage continued observations.


\section{Orbital Dynamics of the Companion}
\label{properties}

The properties of the stellar companion may be further constrained
from orbital dynamics considerations, as we describe in this section.


\subsection{Mass and Orbit}
\label{massorbit}

The mass and separation of the stellar companion to 30~Ari~B may be
constrained from the linear trend detected in the RV data (see Section
\ref{rv}). The trend does not exhibit a ``turn-around'' point where
the slope changes from negative to positive. However, the total
amplitude of the trend over the time baseline of the observations
places a lower limit on the semi-amplitude of the variations due to
the companion. The trend shown in Table \ref{planet} multiplied by the
time baseline (2,536 days) yields a minimum RV amplitude of
$\sim$305~m\,s$^{-1}$.

The left panel of Figure \ref{propfig} shows the resulting
mass/separation limits where the linear trend has been converted to an
acceleration, $\dot{v}$ and then converted to a mass estimate via $M_B
= (\dot{v} a^2) / G$ where $\dot{v} = dv/dt$, $M_B$ is the mass of the
detected binary companion, and we have assumed a circular orbit. The
limit is shown as a solid line and the dashed lines represent the
1$\sigma$ uncertainties propagated from the linear trend
uncertainties. Anything below these lines has either insufficient mass
or proximity to the host star to produce the observed trend. Since we
know from the angular separation (see Section~\ref{detection}) that
the companion semi-major axis is at least 21.9~AU (vertical dotted
line), the companion lower mass limit is $\sim$27 Jupiter masses. The
valid area of parameter-space shown in the figure may thus be
constrained to the shaded region.

An additional constraint on the companion mass may be applied by
extending the above methodology, as described by \citet{how10}. The
physical separation between the 30~Ari~B stars, if they are bound, is
21.9~AU / $\sin(\theta)$ where $\theta$ is the angle between our
line-of-sight and the primary-secondary vector ($\theta=0$ implies the
secondary is behind the primary). If the secondary is the source of
the RV linear trend described in Section \ref{rv}, then it imparts a
line-of-sight acceleration of $\dot{v} = (G M_B / r^2 ) \times
\cos(\theta)$ where $r$ is the physical separation with no assumption
regarding orbital eccentricity. This leads to the following expression
for the companion mass:
\begin{equation}
  M_B = \frac{\dot{v} (21.9 \ \mathrm{AU})^2}{G (1-\cos^2(\theta))
    \cos(\theta)}
\end{equation}
The cubic term in the denominator must be negative since $dv/dt$ is
negative and $M_B$ is positive. Furthermore, because the cosine
function is bound between -1 and 1, the cubic's value lies between
-0.385 and zero. The equation therefore becomes an inequality for
$M_B$:
\begin{equation}
  M_B > \frac{(-0.12 \pm 0.03 \ \mathrm{m/s/day}) \times (21.9 \pm 0.7
    \ \mathrm{AU})^2}{G (-0.385)}
\end{equation}
where we have substituted the linear trend from Table
\ref{planet}. This results in a minimum mass of the stellar companion
of $M_b > 0.29 \pm 0.08 \ M_\odot$, consistent with the companion
being M1-3 as determined in Section \ref{detection}.

The calculated properties of the companion, including the mass of
30~Ari~B and the projected separation, result in a minimum orbital
period of $95 \pm 6$ years. This is consistent with the companion
being of stellar mass producing an observed long timescale RV trend.


\subsection{System Orbital Stability}
\label{stability}

The existence of a planet located $\sim$1~AU from the host star may be
used to place further constraints on the orbit of the stellar
companion. Planets have been detected in both S-type and P-type
orbits, the stability of which have been investigated by numerous
authors \citep{har77,egg95,mus05}. We use the analytical solutions
provided by \citet{hol99} to determine the range of binary separations
and eccentricities that will allow the planetary orbit to remain
stable. To do this, we invert Equation 1 of \citet{hol99} as follows
\begin{eqnarray}
  a_b = a_c / [(0.464 \pm 0.006)] + (-0.380 \pm 0.010) \mu \nonumber \\ +
  (-0.631 \pm 0.034) e + (0.586 \pm 0.061) \mu e \nonumber \\ + (0.150
  \pm 0.041) e^2 + (-0.198 \pm 0.074) \mu e^2]
\end{eqnarray}
where $a_b$ is the binary separation, $e$ is the binary orbital
eccentricity, and $a_c$ is the maximum allowed semi-major axis of the
planet. The mass ratio, $\mu$, is defined as $\mu = m_2 / (m_1 +
m_2)$, and is thus $\mu = 0.5$ for an equal mass binary. The right
panel of Figure \ref{propfig} shows the binary separation and
eccentricity limitations for three different mass ratios with the
constraint that a planet must be allowed to exist at $a_c =
1.01$~AU. We have the additional constraint imposed by the projected
separation of 21.9~AU, represented by the horizontal dotted
line. Based on our spectral type estimate of M1, we adopt a mass for
the companion of 0.5~$M_\odot$ resulting in a mass ratio of $\mu =
0.3$. Thus, the valid regions of the plot exist above both the dotted
and dashed lines. For a companion separation equal to the projected
separation, the eccentricity of the binary orbit must be less than
$\sim 0.75$.


\begin{deluxetable}{lcc}
  \tablecaption{\label{summary} Summary of Stellar Companion Properties}
  \tablehead{
    \colhead{Parameter} &
    \colhead{Value} &
    \colhead{Section}
  }
  \startdata
  Angular Separation (\arcsec)   & $0.536 \pm 0.007$ & \ref{detection} \\
  Projected Separation (AU)      & $21.9 \pm 0.7$    & \ref{detection} \\
  $\Delta J$ magnitude           & $3.15 \pm 0.07$   & \ref{detection} \\
  Apparent $J$ magnitude         & $9.23 \pm 0.07$   & \ref{detection} \\
  Absolute $J$ magnitude         & $6.18 \pm 0.09$   & \ref{detection} \\
  Spectral type                  & M1-3              & \ref{detection} \\
  RV linear trend (m/s/day)      & $-0.12 \pm 0.03$  & \ref{rv} \\
  Mass ($M_\odot$)               & $> 0.29 \pm 0.08$ & \ref{massorbit} \\
  Orbital Period (years)         & $< 95 \pm 6$      & \ref{massorbit} \\
  Orbital Eccentricity ($a_b = 21.9$~AU) & $< 0.75$  & \ref{stability}
  \enddata
\end{deluxetable}

\section{Conclusions}
\label{con}

In this paper, we have presented significant new observations that
attempt to describe the detected objects orbiting 30~Ari~B. Table
\ref{summary} summarizes our derived parameters of the stellar
companion to 30~Ari~B. The 30~Ari system as a whole is clearly quite
complex with the A and B components harboring planetary and stellar
companions. This complexity may be attributed partially to the
relative youth of the system since A and B are each less than 1 Gyr in
age \citep{gue09}, although the hierarchical structure of the system
is likely stable for long timescales. Additionally, the relatively
large (minimum) separation of the detected stellar companion to
30~Ari~B produces orbital motion that makes it difficult to constraint
the orbital inclination. If the companion and the known planet are
coplanar then that would have significant implcations for the
formation and evolutionary history of the system and provide
additional constraints on the overall system stability. Further
observations of the companion will be able to improve our knowledge of
the inclination and the kinematics of the system.

The rather unusual nature of the system as described raises the issue
of appropriate system component names. The reader will have noticed
that we have thus far avoided assigning a name to the companion. The
nomenclature of such systems is as complex as the system itself, an
example of which is described by \citet{wri13}. One possibility, that
uses binary star and exoplanet naming conventions, would be to rename
the primary and secondary components of 30~Ari~B to 30~Ari~BA and
30~Ari~BB respectively, leading to a corrected name for the planetary
companion of 30~Ari~BA~b. The guidelines of the Washington
Multiplicity Catalog standard \citep{rag10} recommends that the
stellar components of 30~Ari~B be named 30~Ari~Ba and 30~Ari~Bb,
leading to a collision with the planet naming convention. A compromise
would be to name the newly-detected companion 30~Ari~C (also advocated
by \citet{rob15}), allowing the planet to remain as 30~Ari~Bb. This
would avoid having a name-change for the planet change, which is
desirable from a literature paper-trail perspective. We propose to
adopt this latter as a provisional naming convention for the system,
as also adopted by \citet{rid15}. As it seems that many of the
exoplanet host stars are part of binary systems, we can look forward
to further discussion and adjustment of names and orbital parameters
in the years ahead.


\section*{Acknowledgements}

The authors would like to thank Elliott Horch, Steve Howell, and
Suvrath Mahadevan for several useful discussions. Thanks are also due
to the anonymous referee whose helpful comments improved the
manuscript. D. Huber acknowledges support by the Australian Research
Council's Discovery Projects funding scheme (project number
DE140101364) and by NASA under Grant NNX14AB92G issued through the
Kepler Participating Scientist Program. E.V. Quintana is supported by
a NASA Senior Fellowship at the Ames Research Center, administered by
Oak Ridge Associated Universities through a contract with NASA. This
work made use of the Digitized Sky Survey (DSS) hosted by the Mikulski
Archive for Space Telescopes (MAST). This work was supported by a NASA
Keck PI Data Award, administered by the NASA Exoplanet Science
Institute. Data presented herein were obtained at the W. M. Keck
Observatory from telescope time allocated to the National Aeronautics
and Space Administration through the agency’s scientific partnership
with the California Institute of Technology and the University of
California. The Observatory was made possible by the generous
financial support of the W. M. Keck Foundation. The authors wish to
recognize and acknowledge the very significant cultural role and
reverence that the summit of Mauna Kea has always had within the
indigenous Hawaiian community. We are most fortunate to have the
opportunity to conduct observations from this mountain.


\end{document}